\documentclass[12pt,rotate,epsf]{article}
\usepackage{rotate}
\usepackage{epsfig,epsf}
\advance \textheight by \topskip
\def\spose#1{\hbox to 0pt{#1\hss}}
\def\simlt{\mathrel{\spose{\lower 3pt\hbox{$\mathchar"218$}}
     \raise 2.0pt\hbox{$\mathchar"13C$}}}
\def\simgt{\mathrel{\spose{\lower 3pt\hbox{$\mathchar"218$}}
     \raise 2.0pt\hbox{$\mathchar"13E$}}}
\def\simpropto{\mathrel{\spose{\lower 3pt\hbox{$\mathchar"218$}}
     \raise 2.0pt\hbox{$\propto$}}}

\newcommand{\R}{{\cal R}}
\newcommand{\Rns}{{\cal R}_{ NS}}

\newcommand{\NUlim}{\nu_{lim}}

\begin{document}

 \begin{center} {\bf \Large 
Orbital Eccentricity Effects on the 
Stochastic Gravitational Wave Background from Coalescing
Binary Neutron Stars}
\bigskip	

{{\bf V.B. Ignatiev$^1$, A.G.Kuranov$^{2}$, K.A. Postnov$^{1,3}$,
M.E. Prokhorov$^2$} \\
$^1$ Faculty of Physics, Moscow State University, 119899 Moscow, Russia\\
$^2$ Sternberg Astronomical Institute,
119899 Moscow, Russia\\
$^3$  Max-Planck Institut f\"ur
Astrophysik, 85740 Garching, Germany}
\end {center}
\bigskip

\begin{abstract}
Unresolved coalescing binary neutron stars in the Galaxy and beyond
produce a  
stochastic gravitational wave background within the LISA frequency 
band, which can be potentially dangerous for possible detection
of fundamentally more interesting relic cosmological backgrounds.
Here we address the question what effects the unavoidable eccentricity
of orbits of these systems should have on the properties of this
backgrounds. In particularly, we examine starting from which frequency 
one-year observations with LISA can be secure from contamination 
by the noise produced by galactic binary neutron stars in eccentric orbits. 
We come to the conclusion that harmonics from 
such binaries do not contribute significantly above $\sim 10^{-3}$
Hz.
\end{abstract}

\section{Introduction}
It is generally recognized that in the 21st century  
gravitational wave astronomy 
will open up a new  window to study physical processes
in strong gravitational field and in the 
early Universe (see Grishchuk et al. 2001 for
a recent review). Among the potential sources of gravitational
waves (GW) for the forthcoming GW detectors, relic stochastic GW 
backgrounds, coming up from the early Universe, play a special role
due to their fundamental importance and 
potential detectability by the space-born laser interferometer
LISA (Grishchuk 1997, Grishchuk et al. 2001). 

The existence of relic gravitational waves is a consequence of quite general
assumptions. In the framework of the conventional cosmological picture, the
strong variable gravitational field of the early Universe amplifies the
inevitable zero-point quantum oscillations of the gravitational waves and
produce a stochastic background of relic gravitational waves in a wide
frequency band from $10^{-18}$~Hz to $10^{10}$~Hz measurable today. The
detection of relic gravitational waves is a feasible way to learn about the
evolution of the very early Universe, up to the limit of Planck era and Big
Bang.

The expected level of this background is model-dependent so its robust
estimation is complicated. Nevertheless, if the level of some background
turns out to be higher than the detector sensitivity limit in a given frequency
band, it can be directly detected using only one interferometer LISA,
provided that other backgrounds do not contribute at these frequencies
(for more deep discussion of the LISA capabilities see recent
papers by Cornish and Larson (2001), Cornish (2001a,b), and
Ungarelli and Vecchio (2001)).

Among various astrophysical GW backgrounds, those produced by several
classes of merging binary compact stars (white dwarfs, neutron stars, black
holes and their combinations) dominate in the LISA frequency band
$10^{-5}-10^{-2}$ Hz 
(e.g. Hils and Bender 1998, Kosenko and Postnov 1999, Ferrari et al. 2001).
A feature of all such backgrounds is that their level (in terms of
the characteristic dimensionless strain metric amplitude $h$) is
$h\propto \sqrt{{\cal R}}$, where $\cal R$ is the merging rate of these
binaries. 
The most "dangerous" from the point of view of "contamination" of the LISA
frequencies are galactic  binary white dwarfs with the highest 
galactic
merging rate ${\cal R}_{wd}\sim 1/300$~yr$^{-1}$.
For one-year observational
run they will make up a stochastic GW background up to frequencies around 1 mHz
(Hils and Bender 1998, Kosenko and Postnov 1999).   

However, if a binary star has a non-circular orbit, it emits GW
at a number of harmonics to its orbital (Keplerian) frequency, 
and above estimates
should be modified. This is precisely the case of 
binary neutron stars, since they generically should reside in
highly eccentric orbits due to two supernova explosions and large
mass loss in the progenitor binary system. 
The possible 
asymmetry of supernova explosion with a natal kick velocity 
of the newborn NS additionally affects the orbital parameters 
by increasing orbital eccentricities.      
Namely, it is not clear from the very beginning 
up to what frequency one can expect such binaries to form a stochastic
background in one year observation (i.e. individual harmonics
cannot be resolved within the frequency resolution bin
$\Delta\nu\sim 3\times 10^{-8}$ Hz). 
The answer to this question is
needed to ensure that indeed there are frequency "windows" inside
the LISA frequency band in which a cosmological GW background can be 
detected.

\section{Gravitational waves in eccentric orbit}

We start with the simple case 
of two point masses in a circular orbit. The energy
losses caused by the quadrapole gravitational wave emission is
\begin{equation}
\label{lo}
L_0= \frac{\displaystyle dE}{\displaystyle dt}(\nu_{K})=
\frac{32}{5} \frac{G^4}{c^5}\frac{m_1^2m_2^2(m_1+m_2)}{a^5}
\end{equation} 
where $G$ is Newton's gravitational constant, $c$ is the speed of light
and $m_1, m_2$ are the masses of stars. All GW radiation in this
case is emitted at the second harmonic to the Keplerian frequency.

For a non-circular orbit with eccentricity $e$, the GW luminosity
increases:
\begin{equation}
\label{le}
\frac{dE}{dt}=
L_0\times\frac{1+\frac{\displaystyle 73}{\displaystyle 24}e^2
+\frac{\displaystyle 37}{\displaystyle 96}e^4}{(1-e^2)^{7/2}}
\end{equation} 
The emission now is widely spread over higher-order harmonics to the 
Keplerian 
frequency  (Peters \& Mathews 1963):
\begin{equation}
\label{le2}
\frac{dE}{dt}(\nu=n\nu_{k})=L_0
\times g(n,e)
\end{equation}
\begin{eqnarray}
\label{g(n,e)}
g(n,e)= \frac{n^4}{32}([J_{n-2}(ne)-2eJ_{n-1}(ne)+\frac{2}{n}J_{n}(ne)\nonumber\\
+2eJ_{n+1}(ne)-J_{n+2}(ne)]^2+(1-e^2)[J_{n-2}(ne)\nonumber\\
-2J_{n}(ne)+J_{n+2}(ne)]^2+\frac{4}{3n^2}{J_{n}}^2(ne))\
\end{eqnarray}
Here $J_n$ is Bessel functions and $n$ is the number of harmonics.
The harmonic at  which the  maximum of energy is radiated 
increases with eccentricity, in accordance with  
the third Kepler's law:
\begin{equation}   
\label{nmax}
n_{max} \approx (1-e^2)^{-3/2}
\end{equation}
This number is asymptotically precise for large eccentricities.        

\section{Merging neutron star binaries}

As we noted above, an important quantity is 
the upper frequency $\NUlim{}$ above which  
each individual source from a given population of binaries 
can be resolved during one-year observation time
(i.e. within the frequency bin $\Delta \nu=3\times 10^{-8}$ Hz). 
In the case of binary white dwarfs, this limiting frequency is of order of 
$\sim 1.2\times 10^{-3}$~Hz for galactic binary WD coalescing rate
$\R=1/300 yr^{-1}$.
The coalescing rate of binary NS 
in most optimistic scenarios is about $\R=10^{-4} yr^{-1}$
so the limiting frequency would be $\NUlim{}=3\times 10^{-4}$~Hz
if they were in circular orbits from the very beginning. However,
it is not the case. 

There is observational evidence for several galactic 
NS binaries (observed as binary
radiopulsars), all of them reciting in eccentric orbits (see the Table).  
The total number of NS binaries in the Galaxy can be roughly estimated 
as $\Rns{}\times T_{gal}= 10^{-4}\times 10^{10}\sim 10^6$,
where $T_{gal}$ is the galactic age. In compact binaries, 
the gravitational radiation back
reaction is the only driving force for orbital evolution.
Given the initial distribution of these system over orbital parameters,
$F_{in}(a,e)$,  
it is straightforward to calculate their steady-state distribution
(see Buitrago et al. 1994
for more detail).

\bigskip
\begin{tabular}{|l|l|l|l|}
\multicolumn{4}{c}{\bf Binary PSR with NS secondaries}\\
\hline
PSR&Period(days)&eccentricity,e&$t_{coalescing},yr$\\
\hline
\hline
B1913+16&0.323&0.617&$1.0\times 10^8$\\
\hline
B1534+12&0.420&0.274&$1.0\times 10^9$\\
\hline
B2721+11c&0.335&0.681&$8.0\times 10^7$\\
\hline
B2303+46&12.340&0.658&$\>1.6\times 10^{12}$\\
\hline
\end{tabular}

\section{Delta-function examples}

\begin{figure}
\centerline{\epsfysize=0.7\hsize 
\rotate[r]{
\epsfbox{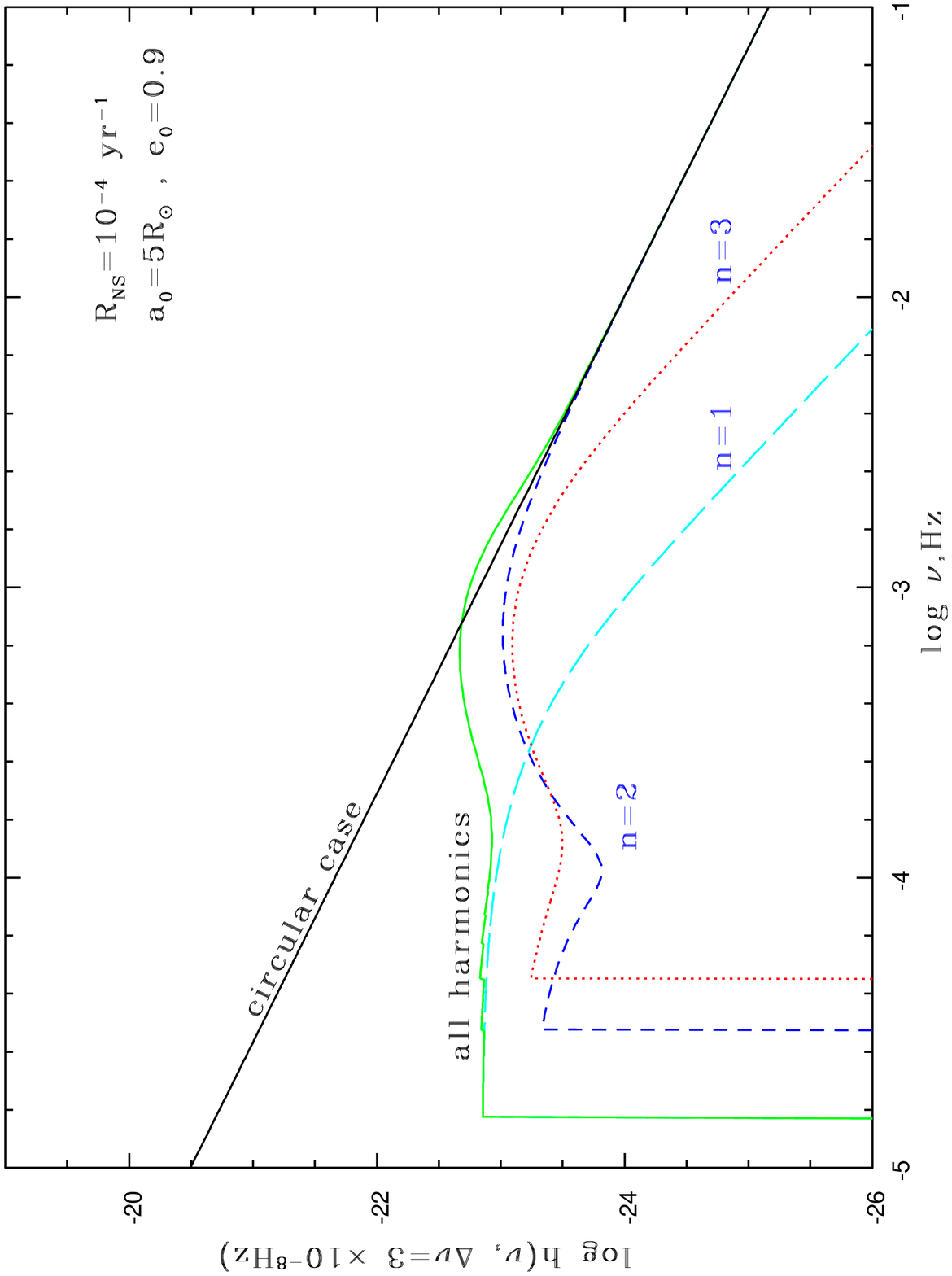}}}
\caption{The GW background from a model 
stationary population of binary NS with initial delta-function like
distribution at $a_0=5 R_{\odot},e_0=0.9$ 
and the contribution from the first, second and third  harmonics.
For comparison, the would-be background from initially circual NS binaries
coalescing with the same rate is shown.
}
\label{delta1}
\bigskip
\centerline{\epsfysize=0.7\hsize 
\rotate[r]{
\epsfbox{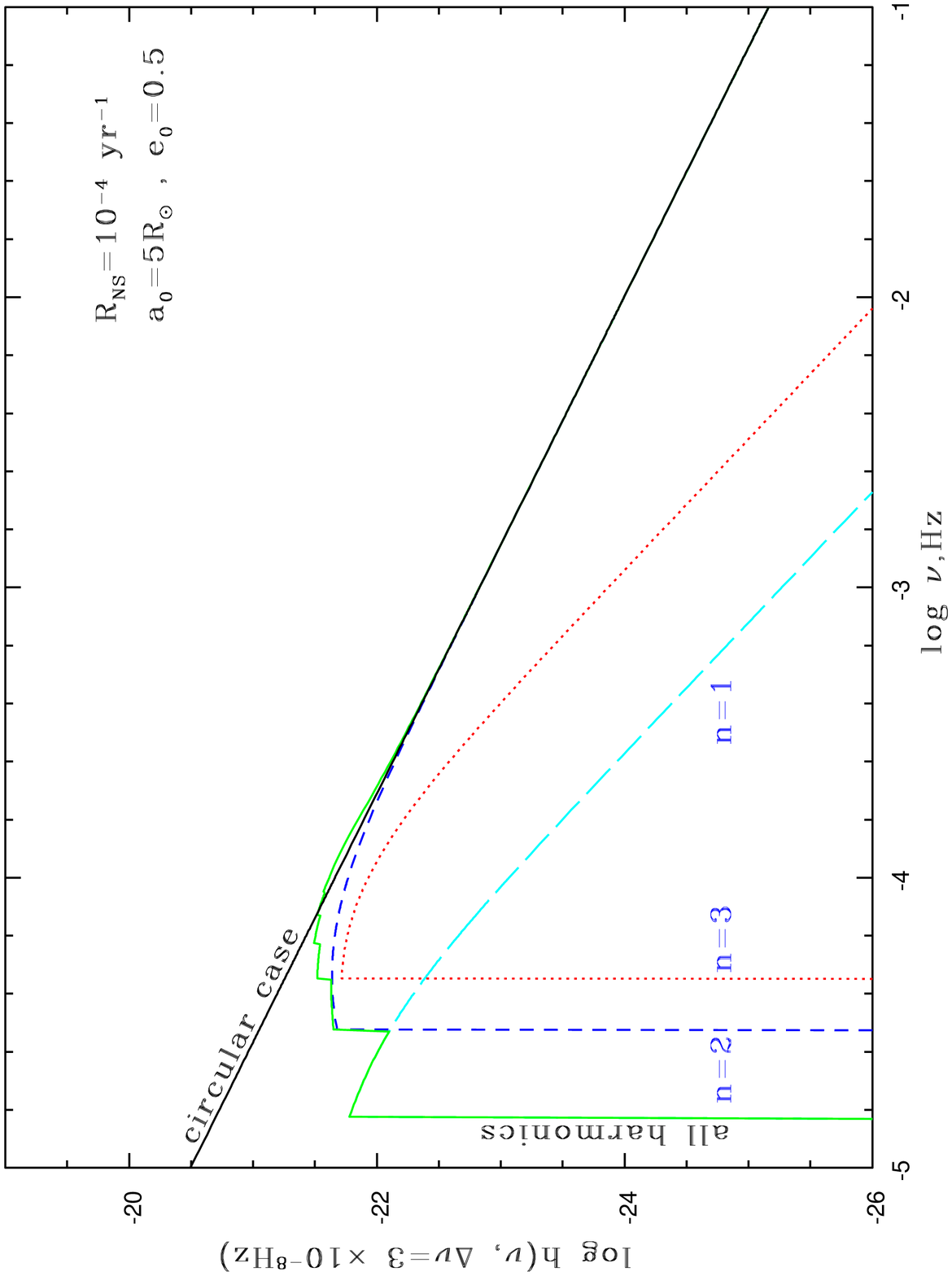}}}
\caption{The same as in Fig~\ref{delta1} for 
systems with initial $a_0=5 R_{\odot},e_0=0.5$. }
\label{delta2}
\end{figure}

Some features of interest here can be learned already
from an (un realistic) 
delta-function-like initial distribution of systems:
$F_{in}(a,e)=\delta(a-a_0,e-e_0)$ (hereafter we assume 
the NS+NS coalescing rate to be 
$\Rns{}=10^{-4}$ yr$^{-1}$). 
The stationary stochastic GW background 
calculated is shown in
Fig.~\ref{delta1} and Fig.~\ref{delta2} 
for $a_0=5 R_\odot$ and $e_0=0.5,~0.9$, respectively. 
The contribution of first three harmonics is shown by separate curves.
It is seen that 
increase in the initial eccentricity strongly affects the shape
of the background up to some frequency at which the eccentricities of 
binaries in the stationary distribution become sufficiently small. 
Above this frequency only the second harmonics from almost circular 
binaries contributes to the total signal. 
The non-monotonic dependence of the background is
due to the non-monotonic behavior of the energy emitted at every harmonics 
with eccentricity.

\section{Stationary stochastic GW noise from coalescing binary NS}
 
Now we turn to calculation of a realistic 
GW background produced by binary NS. This can be done with the following
steps.

(1) Calculate the 
initial galactic distribution of binary NS stars over
orbital semi-major axes and eccentricities $F_{in}(a,e)$
using e.g. binary population
synthesis method (Lipunov et al. 1996). 

(2) Calculate a stationary distribution
function $F_{st}(a,e)$  (Buitrago et al. 1994, Pierro V. and Pinto I.M 1996). 

(3) Compute the total stochastic GW background from these sources.

We shall
assume for simplicity that 
all the sources are at one and the same distance $r=7.9$~kpc away, 
which is close to
the average distance to stars in our Galaxy. This simplifying assumption
does not change our general conclusions significantly.
Basically, at each frequency $\nu$ 
we sum up the GW flux from all the harmonics that fall 
within the chosen frequency bin $\Delta\nu=3\times 10^{-8}$~Hz from 
lower-frequency non-circular systems in the calculated stationary distribution
$F_{st}(\nu,e)$.

\begin{figure}[h]
\centerline{\epsfysize=0.7\hsize 
\rotate[r]{
\epsfbox{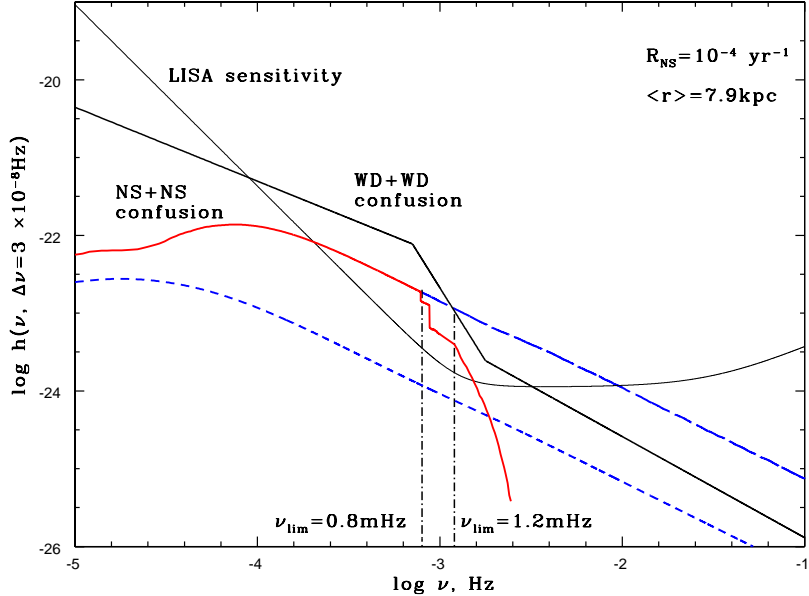}}}
\caption{
GW background from a stationary binary NS population  
calculated for a model spiral 
galaxy with the NS+NS coalescing rate ${\cal R}_{NS}=10^{-4}$~yr$^{-1}$
(the middle broken curve). The limiting frequency above which 
the most powerful harmonics can be resolved in one-year observation
is  $\NUlim{}\simeq 0.8$~mHz. If no harmonics were subtracted from the
frequency bin $3\times 10^{-8}$~Hz, 
the background would continue toward higher frequencies 
as shown by the long-dashed line. The lower dashed line shows
contribution from extragalactic NS+NS binaries.
The upper solid line schematically represents WD+WD confusion limit 
from Bender and Hils (1997) ($\NUlim{}\simeq 1.2$~mHz, Eq. (1)). 
The proposed LISA sensitivity 
is from Thorne (1995) [Fig. 14].}
\label{spectr}
\end{figure}

\begin{equation}
\begin{array}{r}
\label{de/dt/dnu}
\frac{\displaystyle dE}{\displaystyle dt} 
=\int\limits_{\nu}^{\nu+\Delta\nu}\biggl(
~\sum\limits_{n=1}^\infty \int\limits_0^1
L_0(\frac{\displaystyle \nu}{\displaystyle n})
\, g(n,e)
\,F_{st}(\frac{\displaystyle \nu}{\displaystyle n},e)\, de
\biggr)\,\frac{\displaystyle d\nu}{\displaystyle n}
\\
\approx\biggl(~\sum\limits_{n=1}^{n_{lim}} \int\limits_0^1
L_0(\frac{\displaystyle \nu}{\displaystyle n})
\,g(n,e)
\,F_{st}(\frac{\displaystyle \nu}{\displaystyle n},e)\, de\biggr)
\frac{\displaystyle \Delta\nu}{\displaystyle n}
\\
\end{array}
\end{equation}
In the last equation we take into account $\Delta \nu \ll \nu$.
The amplitude of high-order harmonics rapidly decreases, so we stopped 
the summation of harmonics for $n>n_{lim}$, 
$n_{lim}: g(n_{lim},e)= \epsilon \times \max_n[g(n,e)]$.
We assumed $\epsilon=10^{-4}, 10^{-6}, 10^{-10}$. Decreasing 
$\epsilon$ increases the total number of harmonics which 
contribute to the given frequency bin, but practically 
does not change the number of the most powerful harmonics 
within it (see Fig. \ref{num1}).

The resulting noise curve is shown in Fig. ~\ref{spectr} 
in terms of dimensionless
amplitude $h$ 
\begin{equation}
h^2(\nu)= \frac{G}{c^3 r^2 (\pi \nu)^2} \frac{dE}{dt}
\end{equation} 
As expected, the NS+NS confusion noise lies below WD+WD curve, mainly due to 
lower $\Rns{}$. High-order harmonics from non-circular NS binaries mostly 
contribute at lower frequencies, 
so starting from $\nu\sim 10^{-4}$~Hz the calculated 
noise curve practically coincides with that formed by circular NS binaries 
coalescing with the same rate $\Rns{}=10^{-4}$~yr$^{-1}$. 

Extragalactic NS binaries would also form an isotropic confusion noise,
but the level of extragalactic GW backgrounds is generally 
an order of magnitude lower than the galactic one (e.g. calculations of
Kosenko and Postnov 1998) and is 
beyond reach by the expected LISA sensitivity (the bottom dashed curve in 
Fig. \ref{spectr}).
The limiting frequency $\NUlim{}$ for extragalactic NS binaries finds from
the condition for circular systems and can be as higher as $\sim 0.3$ Hz 
(Ungarelli and Vecchio 2000).

\section{The limiting frequency}

\begin{figure}
\centerline{\epsfysize=0.7\hsize 
\rotate[r]{
\epsfbox{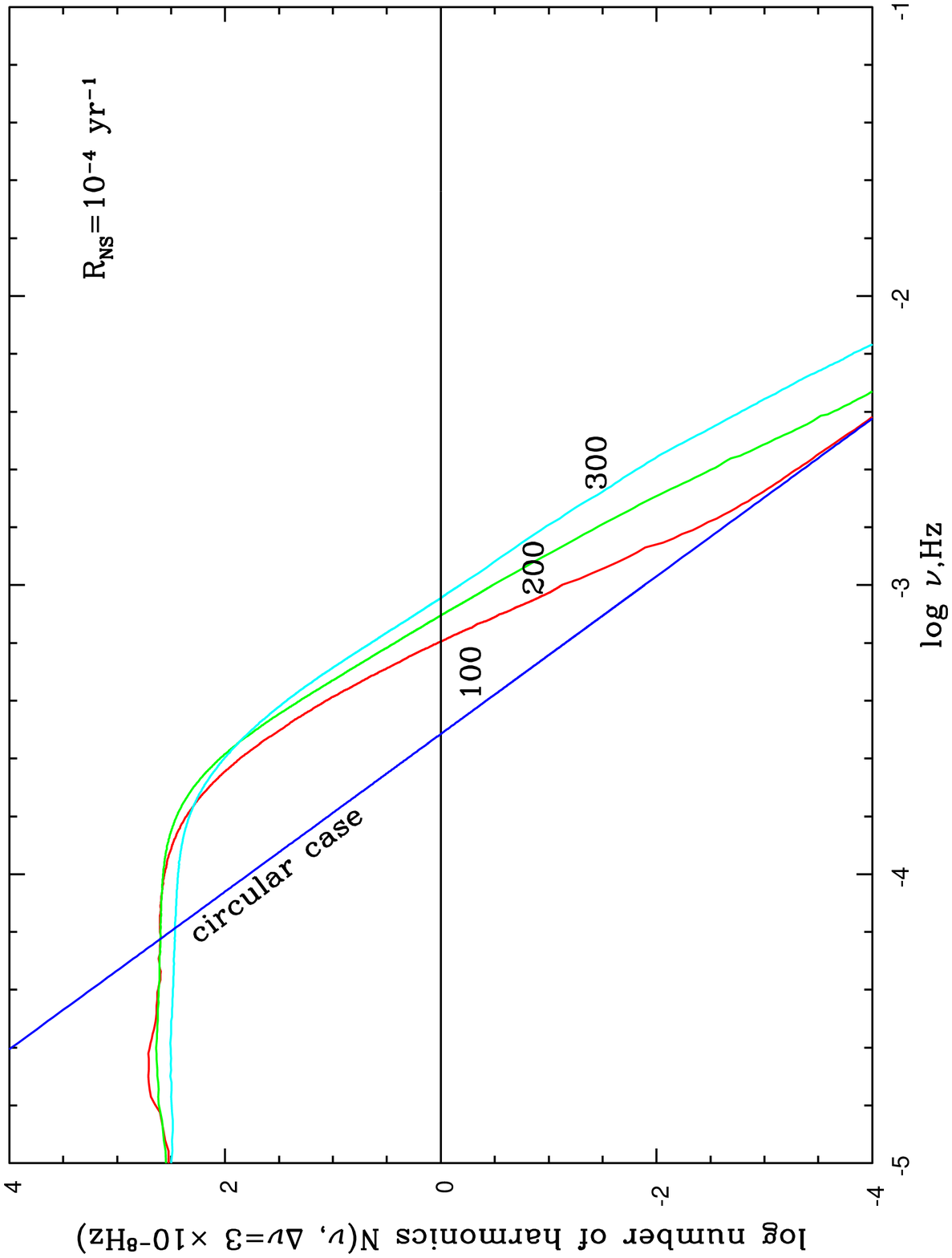}}}
\caption{Number of harmonics from non-circular galactic NS binaries
producing the stochastic noise 
in the frequency bin $\Delta \nu=3\times10^{-8}$~Hz as a function of
the neutron star kick velocity. The curves are labeled with
kick velocity amplitudes $100, 200$ and $300$ km/s assumed in 
the population synthesis calculations.
The straight line is for the circular NS binaries.}
\label{num}
\bigskip
\centerline{\epsfysize=0.7\hsize 
\rotate[r]{
\epsfbox{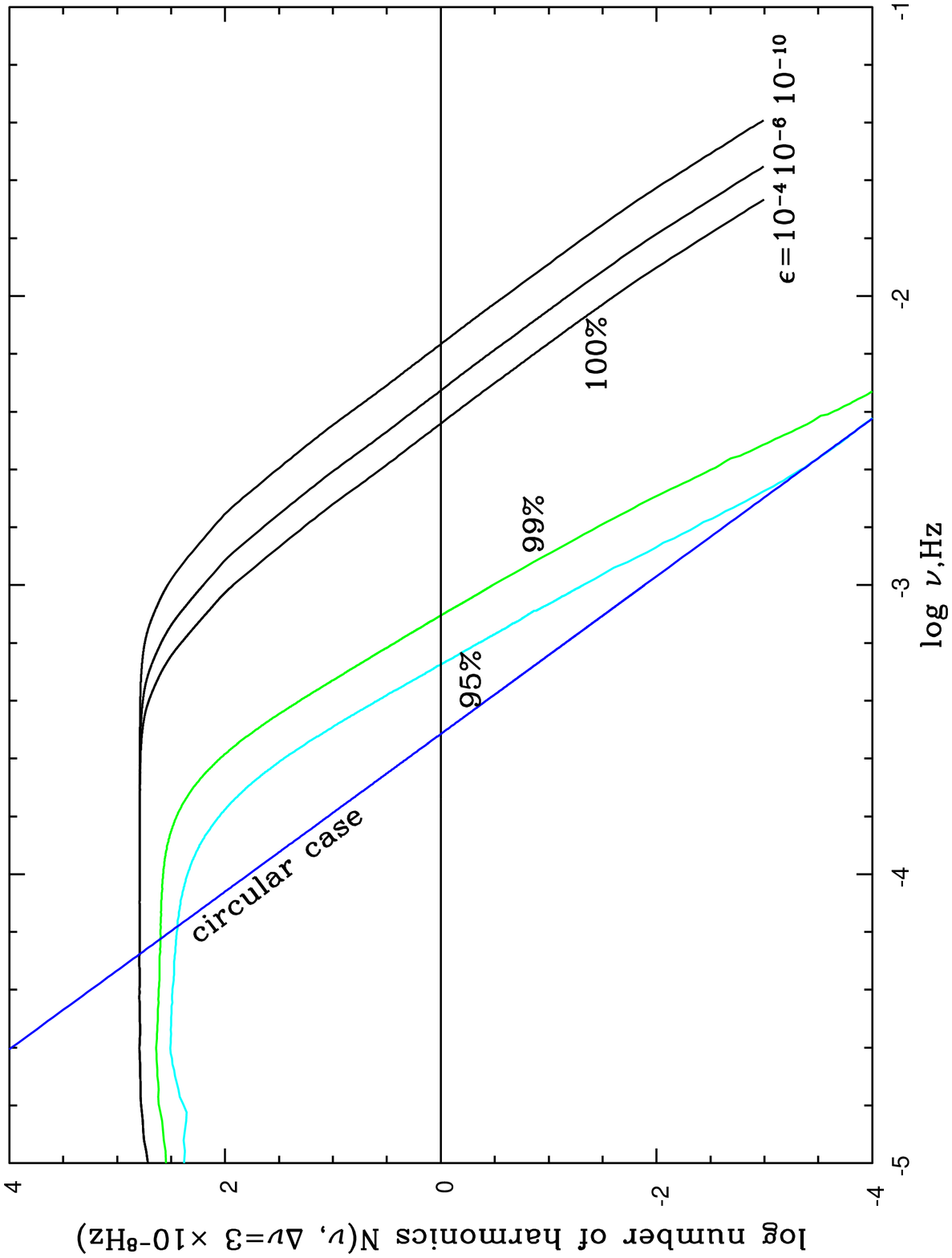}}} 
\caption{Number of the most powerful 
harmonics from non-circular galactic NS binaries which contribute 
95\%, 99\% and 100\% of the total GW energy 
collected in the bin $\Delta\nu=3\times 10^{-8}$
Hz. The effect of 
increase in the number of harmonics $n_{lim}$ determining
the total energy collected in the bin is shown 
for different $\epsilon$. 
Kick velocity amplitude $200$ km/s.}
\label{num1}
\end{figure}

More important in the non-circular case is the increase of the limiting 
frequency $\NUlim{}$ above which individual 
sources can be resolved during a one-year observation.
Formally, the number of harmonics inside the frequency bin $\Delta\nu$.  
is more than one at any frequency,
and in this sense the 
GW background extends up to very high frequency $\sim 1KHz$.
But at high frequencies, the main contribution must come
from circular binaries (2-nd harmonics), with the total GW power 
from higher-order harmonics from low-frequency eccentric 
systems becoming gradually less and less.  
So the number of harmonics 
contributing in a given frequency bin $\Delta\nu=3\times 10^{-8}$~Hz
were counted starting with 
the strongest one and continuing until 99
bin~(Eq.(\ref{de/dt/dnu})) had been included.
The number of such harmonics as a function of frequency and the
assumed neutron star kick velocity amplitude (which mostly
affect the population synthesis results) 
$N(\nu,\Delta\nu)$ is shown in Fig.~\ref{num}. The limiting frequency
$\NUlim{}$ is determined from equation $N(\NUlim{},\Delta\nu)=1$.
To within uncertainties of our calculations ($\Rns{},F_{in}(a,e)$,
etc.), 
$\NUlim{}\simeq 10^{-3}$~Hz, close to the break in the
confusion limit from binary WD. 
For comparison, in the same Fig.~\ref{num} we
show $N(\nu,\Delta\nu)$ for stationary circular NS binaries coalescing with
the same rate $\Rns{}=10^{-4}$~yr$^{-1}$ 
($\NUlim{}({e=0})\propto 3\times 10^{-4}$~Hz).
The effect of increasing  the chosen level for the GW power
from the strongest harmonics in the bin is 
illustrated in Fig. \ref{num1}. Increasing this level from 95\%
to 100\% 
changes the  limiting frequency  $\NUlim{}$
by almost an order of magnitude.  

This procedure, however, is not complete. Indeed, consider a frequency bin
next to thus defined $\nu_{lim}$. We may ask the question which harmonics
will be most probably absent inside it. The answer is those which are the
least probable at this frequency. The probability to find a harmonics in the
bin at a given frequency is determined by the number of the harmonic and the
stationary distribution function of sources $F_{st}(\nu,e)$. It happens that
this is harmonics number one (mostly due to a steep decrease of the systems'
stationary distribution function). If we remove all 1st harmonics of systems
with orbital frequencies falling into the chosen bin from the
sum~Eq.(\ref{de/dt/dnu}), we are left again with some (smaller) GW power in
the bin and may wish to find the limiting frequency exactly in the same way
as described above (i.e. by fixing some level and summing up the strongest
harmonics up to this level), call it $\nu_{lim,-1}$. Above this new limiting
frequency, we can repeat the entire procedure to find $\nu_{lim, -2}$ (this
time the 2d harmonics is least probable to be found in the bin next to the
new limiting frequency), etc., until the noise level of the detector is
reached. The step-like line which continues the spectrum above $\sim 1$ mHz
in Fig. \ref{spectr} illustrates this procedure. This line of course do not
represent the "real" GW noise curve from binary NS and just gives a feeling
of how it most probably behaves at $\nu_{lim}$, above which individual
harmonics from coalescing binary NS can be singled out.

\begin{figure}[h]
\centerline{\epsfysize=0.7\hsize 
\rotate[r]{
\epsfbox{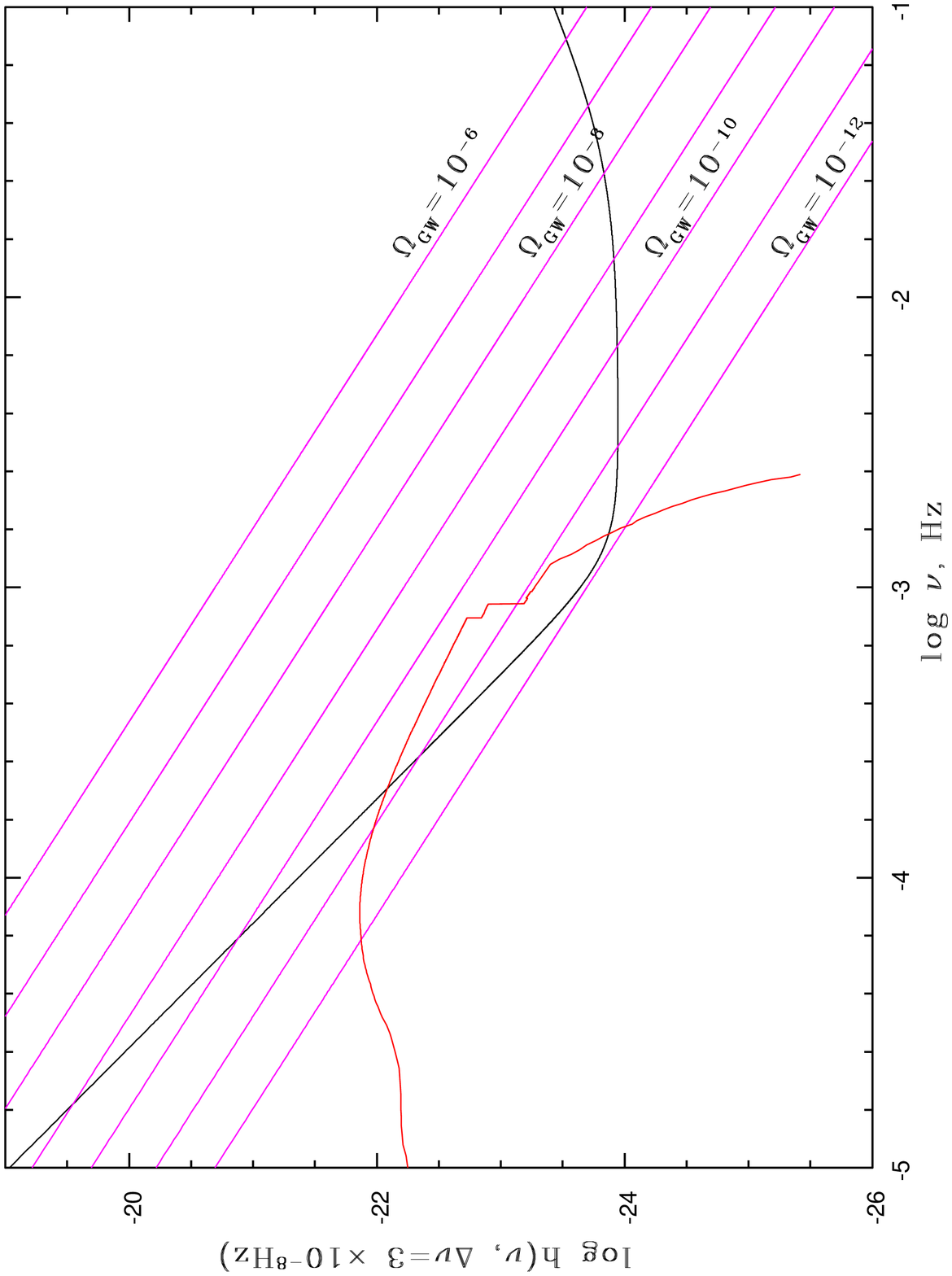}}}
\caption{Levels of equal $\Omega_{GW}$ in comparison with 
the expected stationary galactic NS+NS 
stochastic background in the LISA frequency band.}
\label{spectr_brane}
\end{figure} 

\section{Discussion and conclusions}

A stochastic GW background is frequently described by the quantity
$\Omega_{GW}(\nu)$, defined as the ratio of the
energy density GW in the bandwidth equal to
frequency to the total energy density to close the Universe 
(see critical discussion of this quantity in Grishchuk et al. 2001).
In terms of this quantity, any stochastic cosmological GW
background  with   
$\Omega_{GW} \simeq 10^{-11} \div (\hbox{a few}) \times 10^{-12}$
can be detected by LISA interferometer within the frequency
range from $\sim 10^{-3}$ to $\sim 10^{-2}$ Hz (see Fig. (\ref{spectr_brane})).

The primordial nucleosynthesis arguments give an upper limit for possible
cosmological GW density $\Omega_{GW} \le 10^{-6}$. The so-called "Standard
Inflationary Model" predicts the energy density of relic gravitational waves
at an even smaller level
$\Omega_{GW} \approx (a\,few)\times 10^{-14}$. So weak a background cannot
be certainly detected by LISA. However, more realistic appears the model
(Grishchuk et al. 2001) in which the cosmological stochastic gravitational
wave background level can be well above the LISA sensitivity limit.
Detecting such a background by the LISA interferometer remains one of the
most exciting goals of this challenging project.
\vspace{1cm}

{\it Acknowledgments}. The authors are pleased to 
acknowledge the organizers of the Baksan School for 
financial support. The work was also partially 
supported by RFBR grants
00-02-17164 and 99-02-17884-a. 


{\small
\vskip 10pt
\vskip 6pt
\begin{thebibliography}{30}
\bibitem{}
Bender, P.~L., Hils, D.,~1997, Class. Quantum. Grav., \textbf{14}, 1439
\bibitem{}
Bender, P.~L. et al.,~2000, LISA: Laser Interferometer Space Antenna. 
A Cornerstone Mission for the observation of gravitational waves -- System 
and Technology Study Report, ESA-SCI(2000)11; available at ftp://ftp.rzg.mpg.de/
pub/grav/lisa/sts-1.04.pdf
\bibitem{}
Buitrago, J., Moreno--Garrido, C., and Mediavilla, E., 1994, MNRAS, \textbf{268}, 841
\bibitem{}
Einstein, A.~1916, Preuss. Akad. Wiss. Berlin, Sitzungsberichte der 
physikalischmathematischen Klasse, \textbf{1}, 688
\bibitem{}
Einstein, A.~1918, Preuss. Akad. Wiss. Berlin, Sitzungsberichte der 
physikalischmathematischen Klasse, \textbf{1}, 154 
\bibitem{}
Ferrari, V., Matarrese, S., and Schneider, R.~1999, MNRAS, \textbf{303}, 258
\bibitem{}
Cornish N.J.  gr-qc/0106058 
\bibitem{}
Cornish N.J.  astro-ph/0105374
\bibitem{}
Cornish N.J., Larson S.L. gr-qc/0103075
\bibitem{}
Grishchuk, L.~P., Lipunov, V.~M., Postnov, K.~A., Prokhorov, M.~E.,
and Sathyaprakash, B.~S.,
2001, Usp. Fiz. Nauk, \textbf{171}, 3
\bibitem{}
Kosenko D.~I., Postnov, K.~A.,~1998, Astron. Astophys., \textbf{336}, 789K
\bibitem{}
Landau, L.~D., Lifshiz, E.~M.~1971, Classical Theory of Fields, 
Addisson-Wesley, Reading, Massachusetts and Pergamon Press, London
\bibitem{}
Lipunov, V.~M., and Postnov, K.~A.,~1987, Soviet Astron., \textbf{31}, 228
\bibitem{}
Lipunov, V.~M., Postnov, K.~A., and Prokhorov M.~E.,~1996, Astron. Astophys., \textbf{310}, 489
\bibitem{}
Lipunov, V.~M., Postnov, K.~A., and Prokhorov, M.~E.,~1997, MNRAS, \textbf{288}, 245
\bibitem{}
Peters, P.~S., and Mathews, J.,~1963, Phys. Rev., \textbf{131}, 435
\bibitem{}
Pierro V., Pinto I.M  ApJ, 469, 272, 1996
\bibitem{}
Postnov K.A., Prokhorov M.E.,~1997, Astron. Astophys., \textbf{327}, 428
ApJ, 494, 674 
\bibitem{}
Thorne. K.~S,~1995, in \emph{Particle and Nuclear Astrophysics and Cosmology in the Next Millennium},
edited by E.~W.~Kolb, and R.~D.~Peccei, World Scientific, Singapore, 1995, p.160.
\bibitem{}
Ungarelli C., and Vecchio A.,~2000, gr-gc/0003021.
\bibitem{}
Ungarelli C., and Vecchio A.,~2001, astro-ph/0106538.
\bibitem{}
Vecchio A.,~1997, Class. Quantum. Grav., \textbf{14}, 1431 


\end{thebibliography}
}
\end{document}